## Title

- Enhanced superconductivity in aluminum-based hyperbolic metamaterials

## Authors

V. N. Smolyaninova,[1] C. Jensen,[1] W. Zimmerman,[1] J. C. Prestigiacomo,[2] M. S. Osofsky,[2] H. Kim,[2] Z. Xing,[3] M. M. Qazilbash,[3] I. I. Smolyaninov[4]*

## Affiliations

[1]Department of Physics Astronomy and Geosciences, Towson University, 8000 York Rd., Towson, MD 21252, USA.
[2]Naval Research Laboratory, Washington, DC 20375, USA.
[3]Department of Physics, College of William and Mary, Williamsburg, Virginia 23187-8795, USA.
[4]Department of Electrical and Computer Engineering, University of Maryland, College Park, MD 20742, USA.
*Corresponding author. E-mail: smoly@umd.edu

## Abstract

One of the most important goals of condensed matter physics is materials by design, i.e. the ability to reliably predict and design materials with a set of desired properties. A striking example is the deterministic enhancement of the superconducting properties of materials. Recent experiments have demonstrated that the metamaterial approach is capable of achieving this goal, such as tripling the critical temperature $T_c$ in Al-$Al_2O_3$ epsilon near zero (ENZ) core-shell metamaterial superconductors. However, transport properties of such metamaterials remained much worse compared to conventional superconductors. Here, we demonstrate that an Al/$Al_2O_3$ hyperbolic metamaterial geometry is capable of a similar $T_c$ enhancement, while having superior transport and magnetic properties compared to the core-shell metamaterial superconductors. This result opens up numerous new possibilities for metamaterial enhancement of $T_c$ in other practically important simple superconductors, such as niobium and $MgB_2$. It also indicates that the recently discovered hyperbolic properties of high $T_c$ superconductors may play considerable role in superconductivity enhancement in cuprates.

## MAIN TEXT

## Introduction

One of the most important goals of condensed matter physics is reliably designing new materials with enhanced superconducting properties. Recently, a metamaterial strategy, consisting of deliberately engineering the dielectric properties of a nanostructured "metamaterial superconductor" that results in an enhanced electron pairing interaction that increases the value of the superconducting energy gap and the critical temperature, $T_c$, was suggested to achieve this goal (1, 2). Our recent experimental work (3, 4) has conclusively demonstrated that this approach can indeed be used to increase the critical temperature of a composite superconductor-dielectric metamaterial. For example, we have demonstrated the use of $Al_2O_3$-coated aluminum nanoparticles to form epsilon near zero (ENZ) core-shell metamaterial superconductors with a $T_c$ that is three times that of pure aluminum (4). However, this core-shell metamaterial superconductor geometry exhibits poor transport properties compared to its parent (aluminum) superconductor. A natural way to overcome this issue is the implementation of the hyperbolic metamaterial geometry (Fig. 1A), which has been suggested in (1, 2). Hyperbolic metamaterials are extremely anisotropic uniaxial materials, which behave like a metal ($\mathrm{Re}\varepsilon_{xx} = \mathrm{Re}\varepsilon_{yy} < 0$) in one direction and like a dielectric ($\mathrm{Re}\varepsilon_{zz} > 0$) in the orthogonal direction. Originally introduced to overcome the diffraction limit of optical imaging (5), hyperbolic metamaterials demonstrate a number of novel phenomena resulting from the broadband singular behavior of their density of photonic states. The "layered" hyperbolic metamaterial geometry shown in Fig. 1A is based on parallel periodic layers of metal separated by layers of dielectric. This geometry ensures excellent transport properties in the plane of the layers. As noted in (6), typical high $T_c$ superconductors (such as BSCCO) exhibit hyperbolic behavior in a substantial portion of the far infrared and THz frequency ranges. In this report, we demonstrate that the artificial hyperbolic metamaterial geometry may also lead to a considerable enhancement of superconducting properties.

Electromagnetic properties are known to play a very important role in the pairing mechanism of superconductors (7). According to the BCS theory, a Cooper pair is formed from two electrons with opposite spins and momenta that are loosely bound. This mechanism may be described as an attractive interaction of electrons that results from the polarization of the ionic lattice which these electrons create as they move through the lattice. Based on this interpretation, Kirzhnits *et al*. formulated their description of superconductivity in terms of the dielectric response function of the superconductor (7). They demonstrated that the electron-electron interaction in a superconductor may be expressed in the form of an effective Coulomb potential

$$V(\vec{q},\omega)=\frac{4\pi e^2}{q^2\varepsilon_{eff}(\vec{q},\omega)}=V_C\frac{1}{\varepsilon_{eff}(\vec{q},\omega)} \tag{1}$$

where $V_C$ is the Fourier-transformed Coulomb potential in vacuum, and $\varepsilon_{\mathrm{eff}}(q,\omega)$ is the linear dielectric response function of the superconductor treated as an effective medium. Following this "macroscopic electrodynamics" formalism, it appears natural to use the recently developed plasmonics (8) and electromagnetic metamaterial (9) tools to engineer and maximize the electron pairing interaction (Eq.1) in an artificial "metamaterial superconductor" via deliberate engineering of its dielectric response function $\varepsilon_{\mathrm{eff}}(q,\omega)$. For example, it was predicted in (1, 2) that considerable enhancement of the attractive electron-electron interaction may be expected in such actively studied metamaterial scenarios as ENZ (10) and hyperbolic metamaterials (5). In both cases $\varepsilon_{\mathrm{eff}}(q,\omega)$ may become small and negative in substantial portions of the relevant four-momentum $(q,\omega)$ space leading to an enhancement of the electron pairing interaction. Indeed, it was demonstrated in (1, 2) that in the case of a hyperbolic metamaterial the effective Coulomb potential from Eq. (1) assumes the form

$$V(\vec{q},\omega)=\frac{4\pi e^2}{q_z^2\varepsilon_2(\vec{q},\omega)+\left(q_x^2+q_y^2\right)\varepsilon_1(\vec{q},\omega)} \tag{2}$$

where $\varepsilon_{xx} = \varepsilon_{yy} = \varepsilon_1$ and $\varepsilon_{zz} = \varepsilon_2$ have opposite signs. As a result, the effective Coulomb interaction of two electrons may become attractive and very strong along spatial directions where

$$q_z^2 \varepsilon_2(\vec{q}, \omega) + \left(q_x^2 + q_y^2\right)\varepsilon_1(\vec{q}, \omega) \approx 0 \tag{3}$$

Demonstration of the resulting superconductivity enhancement in hyperbolic metamaterials would open up numerous new possibilities for metamaterial enhancement of $T_c$ in such practically important simple superconductors as niobium and $MgB_2$.

**Results**

Here, we report the first successful realization of such an artificial hyperbolic metamaterial superconductor, which is made of aluminum films separated by thin layers of $Al_2O_3$. This combination of materials is ideal for the proof of principle experiments because it is easy to controllably grow $Al_2O_3$ on the surface of Al, and because the critical temperature of aluminum is quite low ($T_{c\,Al}$ = 1.2 K (11)), leading to a very large superconducting coherence length $\xi$ = 1600 nm (11). Such a large value of $\xi$ facilitates the metamaterial fabrication requirements since the validity of macroscopic electrodynamic description of the metamaterial superconductor requires that its spatial structural parameters must be much smaller than $\xi$. It appears that the Al/$Al_2O_3$ hyperbolic metamaterial geometry is capable of superconductivity enhancement, which is similar to that observed for a core-shell metamaterial geometry (4), while having much better transport and magnetic properties compared to the core-shell superconductors. The multilayer Al/$Al_2O_3$ hyperbolic metamaterial samples for our experiments (Fig. 2) were prepared using sequential thermal evaporation of thin aluminum films followed by oxidation of the top layer for 1 hour in air at room temperature, as described in the Materials and Methods section below. Here and thereafter a single metamaterial layer is understood as a layer of Al with a layer of $Al_2O_3$ on its top surface.

To demonstrate that our multilayer samples exhibit hyperbolic behavior, we studied their transport and optical properties (Figs. 3-5). The temperature dependences of the sheet resistance of a 16-layer 10nm/layer Al/$Al_2O_3$ hyperbolic metamaterial and a 100 nm thick Al film are shown in Fig. 3A. As illustrated in the logarithmic plot in the inset, the electronic conductivity of the metamaterial approaches conductivity values of bulk aluminum (indicated by the arrow), and is far removed from the parameter space characteristic for granular Al films (12), which is indicated by the gray area in the inset. These results were corroborated by measurements of IR reflectivity of these samples, shown in Fig. 3B. The IR reflectivity of the hyperbolic metamaterial samples was measured in the long wavelength IR (LWIR) (2.5-22.5 μm) range using an FTIR spectrometer, and compared with reflectivity spectra of Al and $Al_2O_3$. While the reflectivity spectrum of bulk Al is almost flat, the spectrum of $Al_2O_3$ exhibits a very sharp step-like behavior around 11 μm that is related to the phonon-polariton resonance, which results from coupling of an infrared photon with an optical phonon of $Al_2O_3$ (13). The step in reflectivity is due to the negative sign of $\varepsilon_{Al2O3}$ near the resonance. The absence of this step in our multilayers indicates that the aluminum layers in our samples are continuous and not intermixed with aluminum oxide. Such a step is clearly observed in reflectivity data obtained from a core-shell Al/$Al_2O_3$ sample shown in Fig. 3B where the aluminum grains are separated from each other by $Al_2O_3$. On the other hand, this step is completely missing in reflectivity spectra of the hyperbolic metamaterial samples (the step at 9 μm observed in the spectrum of a three-layer sample is due to phonon-polariton resonance of the $SiO_2$ substrate). Thus, our transport and optical measurements confirm excellent DC and AC (optical) conductivity of the aluminum layers of the fabricated hyperbolic metamaterials.

The Kramers-Kronig analysis of the FTIR reflectivity spectra of Al and $Al_2O_3$ measured in (4) also allowed us to calculate the $\varepsilon_1$ and $\varepsilon_2$ components of the $Al/Al_2O_3$ layered films in the LWIR spectral range using the Maxwell-Garnett approximation as follows:

$$\varepsilon_1 = n\varepsilon_m + (1-n)\varepsilon_d \tag{4}$$

$$\varepsilon_2 = \frac{\varepsilon_m \varepsilon_d}{(1-n)\varepsilon_m + n\varepsilon_d} \tag{5}$$

where $n$ is the volume fraction of metal, and $\varepsilon_m$ and $\varepsilon_d$ are the dielectric permittivities of the metal and dielectric, respectively (14). Results of these calculations for a multilayer metamaterial consisting of 13 nm thick Al layers separated by 2 nm of $Al_2O_3$ are shown in Fig. 4. The metamaterial appears to be hyperbolic except for a narrow LWIR spectral band between 11 and 18 μm. A good match between the Maxwell-Garnett approximation (Eqs. (4, 5)) and the measured optical properties of the metamaterial is demonstrated by ellipsometry (Fig. 5A) and polarization reflectometry (Fig. 5B) of the samples.

Variable angle spectroscopic ellipsometry with photon energies between 0.6 eV and 6.5 eV on the $Al/Al_2O_3$ metamaterial have been performed using a Woollam Variable Angle Spectroscopic Ellipsometer (W-VASE). For a uniaxial material with optic axis perpendicular to the sample surface and in plane of incidence, ellipsometry provides the pseudo-dielectric function which, in general, depends both on $\varepsilon_1$ and $\varepsilon_2$. However, as demonstrated by Jellison and Baba (15), the pseudo-dielectric function in this measurement geometry is dominated by the in-plane dielectric function $\varepsilon_1$ and is independent of the angle of incidence. We find that the pseudo-dielectric function of the $Al/Al_2O_3$ metamaterial is indeed similar (but not the same) as that of aluminum i.e. metallic as expected from effective medium theory. We also find that the pseudo-dielectric function is rather insensitive to the angle of incidence. The measured results for the real and imaginary parts of the pseudo-dielectric function in Fig. 5A show good agreement with the model for the in-plane dielectric function (Eq. (4)). The calculated data points are based on the real and imaginary parts of $\varepsilon_{Al}$ tabulated in (16). The measured sign of the real part of the pseudo-dielectric function is negative, which suggests metallic in-plane transport. The sign of the real part of $\varepsilon_2$ (and therefore, the hyperbolic character of our samples) was determined by polarization reflectometry, since ellipsometry data are less sensitive to $\varepsilon_2$ (15). Polarization reflectometry also confirmed the negative sign of the real part of $\varepsilon_1$ consistent with ellipsometry data. The extraction of metamaterial parameters from the polarization reflectometry data is described in detail in the Materials and Methods section. The experimentally determined value of $\varepsilon_2$ at 430 nm is $\varepsilon_2 = 1.30+ 0.09i$, and at 600 nm it is $\varepsilon_2 = 1.56+0.21i$.. The positive real component of $\varepsilon_2$ confirms the hyperbolic character of the metamaterial.

The $T_c$ and critical magnetic field, $H_c$, of various samples (Figs. 6,7) were determined via four-point resistivity measurements as a function of temperature and magnetic field, H, using a Physical Property Measurement System (PPMS) by Quantum Design. Even though the lowest achievable temperature with our PPMS system was 1.75 K, which is higher than the critical temperature $T_{cAl}$ = 1.2 K of bulk aluminum, we were able to observe a pronounced effect of the number of layers on $T_c$ of the hyperbolic metamaterial samples. Fig. 6A shows measured resistivity as a function of temperature for the 1-layer, 3-layer and 8-layer samples each having the same 8.5 nm layer thickness. While the superconducting transition in the 1-layer sample was below 1.75 K, and could not be observed, the 3-layer and 8-layer metamaterial samples exhibited progressively higher critical temperature, which strongly indicates the role of hyperbolic geometry in $T_c$ enhancement. A similar set of measurements performed for several samples having 13 nm layer thickness is shown in Fig. 6B.

The measurements of $H_c$ in parallel and perpendicular fields are shown in Fig. 7. Fig. 7A shows measured resistivity as a function of temperature for a 16-layer 13.2 nm layer thickness hyperbolic metamaterial sample. The critical temperature of this sample appears to be $T_c$ = 2.3 K, which is about two times higher than the $T_c$ of bulk aluminum (another transition at $T_c$ = 2.0 K probably arise from one or two decoupled layers or edge shadowing effects where the thickness of the films is not uniform). The inset in Fig. 7A illustrates the measurements of $H_c^{parallel}$ for this sample. The critical field appears to be quite large (~3T), which is similar to the values of $H_c^{parallel}$ observed previously in granular aluminium films (17). However, it is remarkable that such high critical parameters are observed for the films, which are much thicker than granular Al films.

Measurements of the perpendicular critical field $H_c^{perp}$ for the same metamaterial sample, which are shown in Fig. 7B allowed us to evaluate the Pippard coherence length

$$\xi = \sqrt{\frac{\phi_0}{2\pi H_{c2}}} \tag{6}$$

Assuming $H_{c2}^{perp}$ = 100 G (based on the inset in Fig.7B) the corresponding coherence length appears to be $\xi$ = 181 nm, which is much larger than the layer periodicity. Other measured samples also exhibit the coherence length around 200 nm. Therefore, our use of effective medium approach is validated and our multilayer samples should obey the metamaterial theory.

We have also studied changes in $T_c$ as a function of Al layer thickness in a set of several 8-layer Al/$Al_2O_3$ metamaterial samples, as shown in Fig. 8A. The quantitative behaviour of $T_c$ as a function of $n$ may be predicted based on the hyperbolic enhancement of the electron-electron interaction (Eq. (2)) and the density of electronic states, $\nu$, on the Fermi surface which experience this hyperbolic enhancement (see the Materials and Methods section). Fig. 8B demonstrates that the experimentally measured behaviour of $T_c$ as a function of $n$ (which is defined by the Al layer thickness) correlates well with the theoretical fit, which was obtained using Eq. (26) based on the hyperbolic mechanism of $T_c$ enhancement.

## Discussion

The observed combination of transport and critical properties of the Al/$Al_2O_3$ hyperbolic metamaterials is very far removed from the parameter space typical of the granular aluminum films (12, 17). Together with the number of layer and layer thickness dependences of $T_c$ and $H_c$ shown in Figs. 6-8, these observations strongly support the hyperbolic metamaterial mechanism of superconductivity enhancement described by Eqs. (12- 26). The developed technology enables efficient nanofabrication of thick film aluminum-based hyperbolic metamaterial superconductors with a $T_c$ that is two times that of pure aluminum and with excellent transport and magnetic properties. These results open up numerous new possibilities for considerable $T_c$ enhancement in other practically important simple superconductors, such as niobium and $MgB_2$. However, due to their much smaller coherence length (11, 20) metamaterial structuring of these superconductors must be performed on a much more refined scale. The two-fold increase of $T_c$ in an artificial hyperbolic metamaterial superconductor that we have observed suggests that the recently discovered hyperbolic properties of high $T_c$ superconductors (such as BSCCO) (6) may play a considerable role in the high values of $T_c$ observed in cuprates.

## Materials and Methods

### Metamaterial fabrication.

The multilayer $Al/Al_2O_3$ hyperbolic metamaterial samples for our experiments (Fig. 2) were prepared using sequential thermal evaporation of thin aluminum films followed by oxidation of the top layer for one hour in air at room temperature. The first layer of aluminum was evaporated onto a glass slide surface. Upon exposure to ambient conditions a ~ 2 nm thick $Al_2O_3$ layer is known to form on the aluminum film surface (18). Further aluminum oxidation may also be achieved by heating the sample in air. The oxidized aluminum film surface was used as a substrate for the next aluminum layer. This iterative process was used to fabricate thick multilayer (up to 16 metamaterial layers) $Al/Al_2O_3$ hyperbolic metamaterial samples (throughout our paper a single metamaterial layer is understood as a layer of Al with a layer of $Al_2O_3$ on its top surface). Scanning electron microscope images of a cleaved four-layer sample are shown in Fig. 2. This sample consists of four individual 25 nm thick metamaterial layers deposited on top of each other. The first 25 nm thick layer is clearly visible in the bottom image.

### Extraction of metamaterial parameters from the polarization reflectometry data

The metamaterial parameters were extracted from the polarization reflectometry data as described in detail in (19). Reflectivity for s-polarization is given in terms of the incident angle $\theta$ by

$$R_s = \left| \frac{\sin(\theta - \theta_{ts})}{\sin(\theta + \theta_{ts})} \right|^2 , \quad (8)$$

where

$$\theta_{ts} = \arcsin\left( \frac{\sin\theta}{\sqrt{\varepsilon_1}} \right) . \quad (9)$$

Reflectivity for p-polarization is given as

$$R_p = \left| \frac{\varepsilon_1 \tan\theta_{tp} - \tan\theta}{\varepsilon_1 \tan\theta_{tp} + \tan\theta} \right|^2 , \quad (10)$$

where

$$\theta_{tp} = \arctan\sqrt{\frac{\varepsilon_2 \sin^2\theta}{\varepsilon_1\varepsilon_2 - \varepsilon_1 \sin^2\theta}} . \quad (11)$$

We measured p- and s- polarized absolute reflectance on the metamaterial sample using the reflectance mode of the ellipsometer. The reflectance was measured at two photon energies, 2.07 eV (600 nm) and 2.88 eV (430 nm), as shown in Fig. 5B and was normalized to the measured reflectance of a 150 nm gold film. The absolute reflectance of the gold film was obtained from ellipsometry measurements. The estimated uncertainty in the absolute reflectance of the $Al/Al_2O_3$ metamaterial is one percent. In order to obtain the dielectric permittivity $\varepsilon_1$ and $\varepsilon_2$ values, we fit the s- polarized reflectance first, and get the in-plane dielectric function $\varepsilon_1$. We then use the in-plane dielectric function to fit the p-polarized reflectance to obtain the out-of-plane dielectric function, $\varepsilon_2$. The data analysis was done using W-VASE software. At 2.07 eV (600 nm), $\varepsilon_1$ = -7.17+1.86i and $\varepsilon_2$ = 1.56+0.21i, and at 2.88 eV (430 nm), $\varepsilon_1$ = -2.15+0.50i and $\varepsilon_2$ = 1.30+0.09i. It is clear that the real part of the out-of-plane dielectric function is positive while the real part of the in-plane dielectric function is negative, which confirms the dielectric nature along z-axis and metallic nature in the xy-plane i.e. a hyperbolic metamaterial.

**Modeling of the *T*c enhancement in a hyperbolic metamaterial.**

Using Eqs. (4, 5), the effective Coulomb potential from Eq. (2) may be re-written as

$$V(\vec{q},\omega)=\frac{4\pi e^2}{q^2\left(\frac{q_z^2}{q^2}\frac{\varepsilon_d\varepsilon_m}{((1-n)\varepsilon_m+n\varepsilon_d)}+\frac{q_x^2+q_y^2}{q^2}(n\varepsilon_m+(1-n)\varepsilon_d)\right)}\ . \tag{12}$$

Let us assume that the dielectric response function of the metal used to fabricate the hyperbolic metamaterial shown in Fig. 1 may be written as

$$\varepsilon_m(q,\omega)=\left(1-\frac{\omega_p^2}{\omega^2-\omega_p^2q^2/k^2}\right)\left(1-\frac{\Omega_1^2(q)}{\omega^2}\right)...\left(1-\frac{\Omega_n^2(q)}{\omega^2}\right) \tag{13}$$

where $\omega_p$ is the plasma frequency, $k$ is the inverse Thomas-Fermi radius, and $\Omega_n(q)$ are dispersion laws of various phonon modes (20). Zeroes of the dielectric response function $\varepsilon_m(q,\omega)$ of the bulk metal (which correspond to its various bosonic modes) maximize the electron-electron pairing interaction given by Eq. (1). As summarized in (21), the critical temperature of a superconductor is typically calculated as

$$T_c=\theta\ \exp\left(-\frac{1}{\lambda_{eff}}\right), \tag{14}$$

where $\theta$ is the characteristic temperature for a bosonic mode mediating the electron pairing interaction (such as the Debye temperature $\theta_D$ in the standard BCS theory), and $\lambda_{eff}$ is the dimensionless coupling constant defined by $V(q,\omega)=V_C(q)\ \varepsilon^{-1}(q,\omega)$ and the density of states $\nu$ ( see for example (22)):

$$\lambda_{eff}=-\frac{2}{\pi}\nu\int_0^\infty\frac{d\omega}{\omega}\left\langle V_C(q)\,\mathrm{Im}\,\varepsilon^{-1}(\vec{q},\omega)\right\rangle, \tag{15}$$

where $V_C$ is the unscreened Coulomb repulsion. The integral in Eq. (15) is typically simplified to take into account only the contributions from the poles of the inverse dielectric response function $\varepsilon^{-1}(q,\omega)$, while averaging is performed over all the spatial directions.

Let us consider the region of four-momentum $(q,\omega)$ space, where $\omega > \Omega_1(q)$. While $\varepsilon_m = 0$ at $\omega = \Omega_1(q)$, $\varepsilon_m(q,\omega)$ is large in a good metal and negative just above $\Omega_1(q)$. Compared to the bulk metal, the poles of the angular-dependent $\nu V$ product of the hyperbolic metamaterial are observed at shifted positions compared to the zeroes of $\varepsilon_m$, and additional poles may also appear (2). Based on Eq. (12), the differential of the product $\nu V$ may be written as

$$d(\nu V)=\frac{4\pi e^2 n\sin\theta d\theta}{q^2\left(\frac{\varepsilon_d\varepsilon_m}{((1-n)\varepsilon_m+n\varepsilon_d)}\cos^2\theta+(n\varepsilon_m+(1-n)\varepsilon_d)\sin^2\theta\right)}=$$

$$=-\frac{4\pi e^2 ndx}{q^2\left((n\varepsilon_m+(1-n)\varepsilon_d)-\frac{n(1-n)(\varepsilon_m-\varepsilon_d)^2}{(n\varepsilon_d+(1-n)\varepsilon_m)}x^2\right)}\ , \qquad (16)$$

where $x = \cos\theta$, and $\theta$ varies from 0 to π. The latter expression has two poles at

$$\varepsilon_m=\left(\left(1-\frac{1}{2n(1-n)(1-x^2)}\right)\pm\sqrt{\left(1-\frac{1}{2n(1-n)(1-x^2)}\right)^2-1}\right)\varepsilon_d \qquad (17)$$

As the volume fraction, $n$, of metal is varied, one of these poles remains close to $\varepsilon_m = 0$, while the other is observed at larger negative values of $\varepsilon_m$:

$$\varepsilon_m^+\approx -n(1-n)(1-x^2)\varepsilon_d \qquad (18)$$

$$\varepsilon_m^-\approx -\frac{\varepsilon_d}{n(1-n)(1-x^2)} \qquad (19)$$

This situation is similar to calculations of $T_c$ for ENZ metamaterials (23). Since the absolute value of $\varepsilon_m$ is limited (see Eq. (13)), the second pole disappears near $n = 0$ and near $n = 1$. Due to the complicated angular dependences in Eq. (17), it is convenient to reverse the order of integration in Eq. (15), and perform the integration over $d\omega$ first, followed by angular averaging. Following the commonly accepted approach, while integrating over $d\omega$ we take into account only the contributions from the poles given by Eq. (17), and assume the value of $\mathrm{Im}\varepsilon_m = \varepsilon_m''$ to be approximately the same at both poles. The respective contributions of the poles to $d(\nu V)/dx$ may be written as

$$\frac{d(\nu V)}{dx}\approx\frac{2\pi e^2\left((1-n)\left(\left(1-\frac{1}{2n(1-n)(1-x^2)}\right)\pm\sqrt{\left(1-\frac{1}{2n(1-n)(1-x^2)}\right)^2-1}\right)+n\right)}{q^2\varepsilon''_m(1-n)(1-x^2)\sqrt{\left(1-\frac{1}{2n(1-n)(1-x^2)}\right)^2-1}} \qquad (20)$$

Near $n$=0 and $n$=1 these expression may be approximated as

$$\frac{d(\nu V)^+}{dx}\approx\frac{4\pi e^2 n^2}{q^2\varepsilon''_m} \qquad (21)$$

and

$$\frac{d(\nu V)^-}{dx}\approx\frac{4\pi e^2}{q^2\varepsilon''_m(1-x^2)}\ , \qquad (22)$$

respectively. Note that at the $\omega = \Omega_1(q)$ zero of the dielectric response function of the bulk metal the effective Coulomb potential inside the metal may be approximated as

$$V_m(\vec{q},\omega)=\frac{4\pi e^2}{q^2\varepsilon_m(q,\omega)}\approx-\frac{4\pi e^2}{q^2\varepsilon''_m}\ , \qquad (23)$$

so that the coupling constant $\lambda_{\text{eff}}$ of the hyperbolic metamaterial obtained by angular integration of the sum of Eqs. (21) and (22) may be expressed via the coupling constant $\lambda_{\text{m}}$ of the bulk metal:

$$\lambda_{eff} \approx \lambda_m \left( n^2 + \alpha \ln \left| \frac{1+x_0}{1-x_0} \right| \right), \tag{24}$$

where $\alpha$ is a constant of the order of 1 and $x_0$ is defined by the maximum negative value of $\varepsilon_{\text{m}}$, which determines if the second pole (Eq. (19)) exists at a given *n*. Based on Eq.(19),

$$x_0^2 \approx 1 + \frac{\varepsilon_d}{n(1-n)\varepsilon_{m,\max}} \tag{25}$$

If the second pole does not exist then $x_0 = 0$ may be assumed. Based on Eq. (14), the theoretically predicted value of $T_{\text{c}}$ for the hyperbolic metamaterial is calculated as

$$T_c = T_{Cbulk} \exp\left( \frac{1}{\lambda_m} - \frac{1}{\lambda_{eff}} \right) = T_{Cbulk} \exp\left( \frac{1}{\lambda_m} \left( 1 - \frac{1}{n^2 + \alpha \ln \left| \frac{1+x_0}{1-x_0} \right|} \right) \right) \tag{26}$$

assuming the known values $T_{\text{cbulk}} = 1.2$ K and $\lambda_{\text{m}} = 0.17$ for bulk aluminum (23). The predicted behaviour of $T_{\text{c}}$ as a function of *n* is plotted in Fig. 8B showing good agreement with the experiment.

## Acknowledgments

**Funding:** This work was supported in part by NSF grant DMR-1104676.

**Author contributions:** I.S., V.S. , M.O. and M.Q. wrote the main manuscript text, V.S., C.J. and W.Z. fabricated samples, V.S., J.P., M.O., H.K., Z.X., M.Q. and I.S. collected experimental data, I.S. developed theoretical model of hyperbolic metamaterial superconductors. All authors discussed and reviewed the manuscript.

**Competing interests:** The authors declare no competing financial interests.

**Figures and Tables**

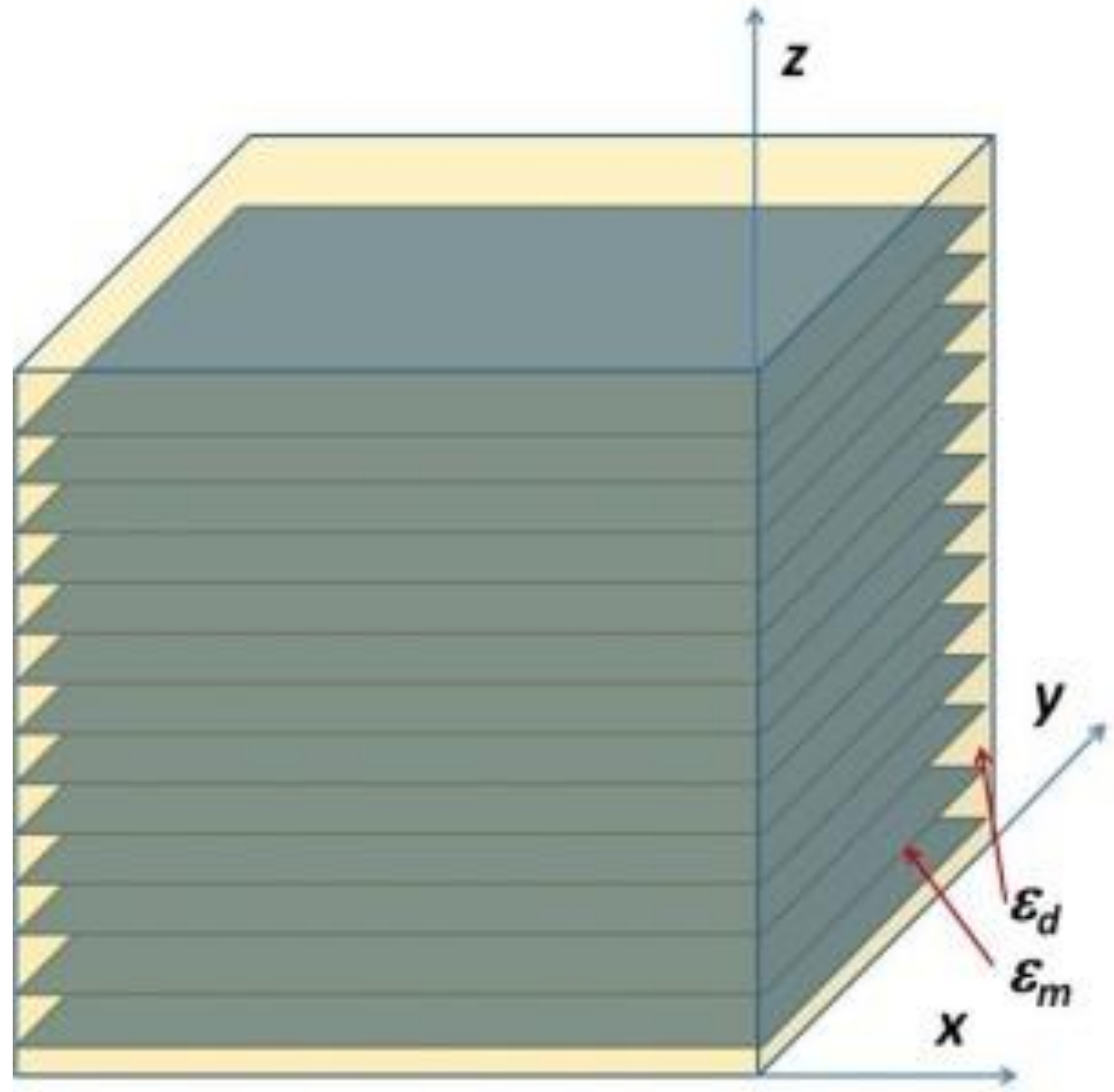

A

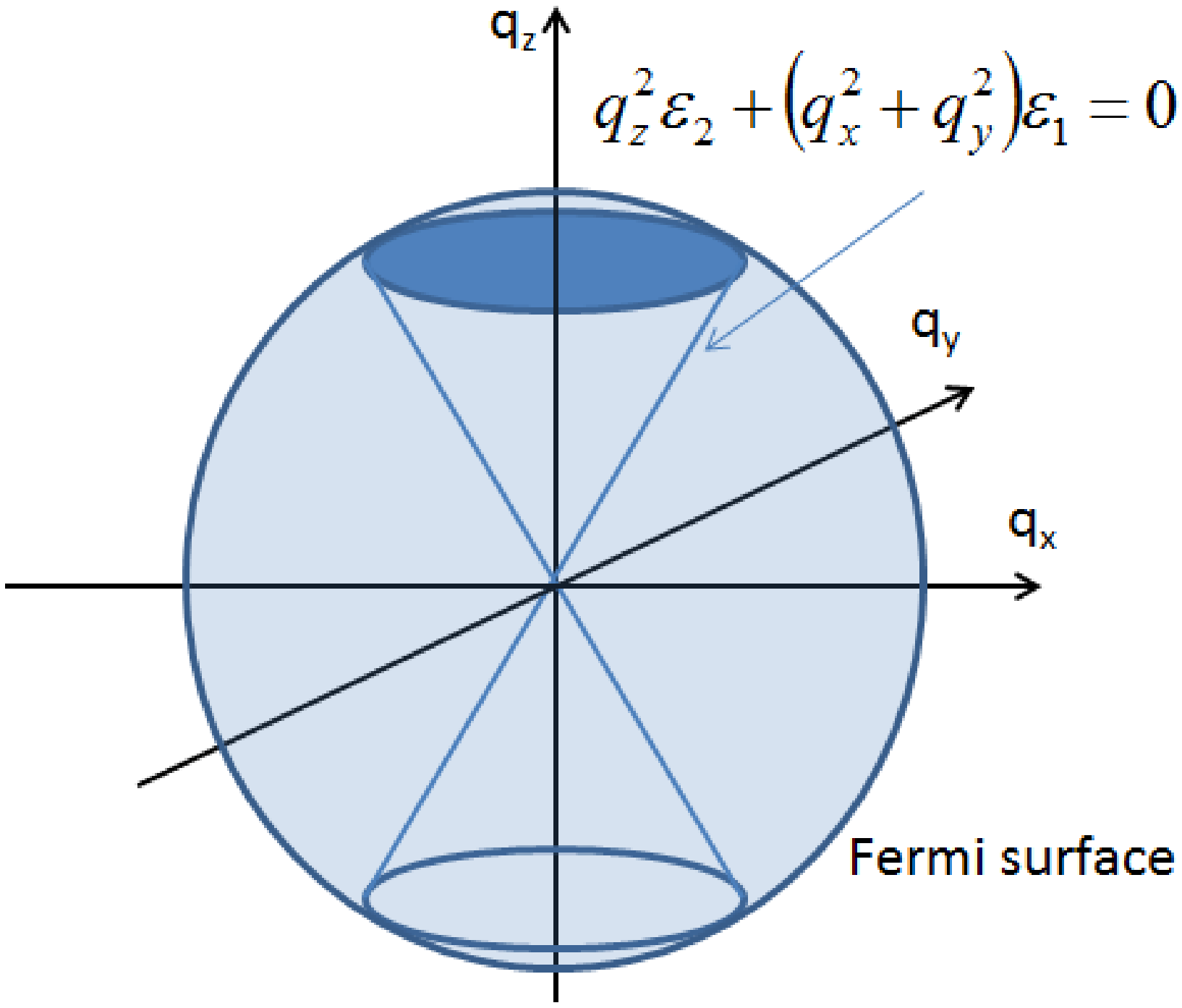

B

**Fig. 1. Geometry and basic properties of hyperbolic metamaterial superconductors.** (**A**) Schematic geometry of a "layered" hyperbolic metamaterial. (**B**) Electron-electron pairing interaction in a hyperbolic metamaterial is strongly enhanced near the cone in momentum space defined as $q_z^2\varepsilon_2 + \left(q_x^2 + q_y^2\right)\varepsilon_1 = 0$.

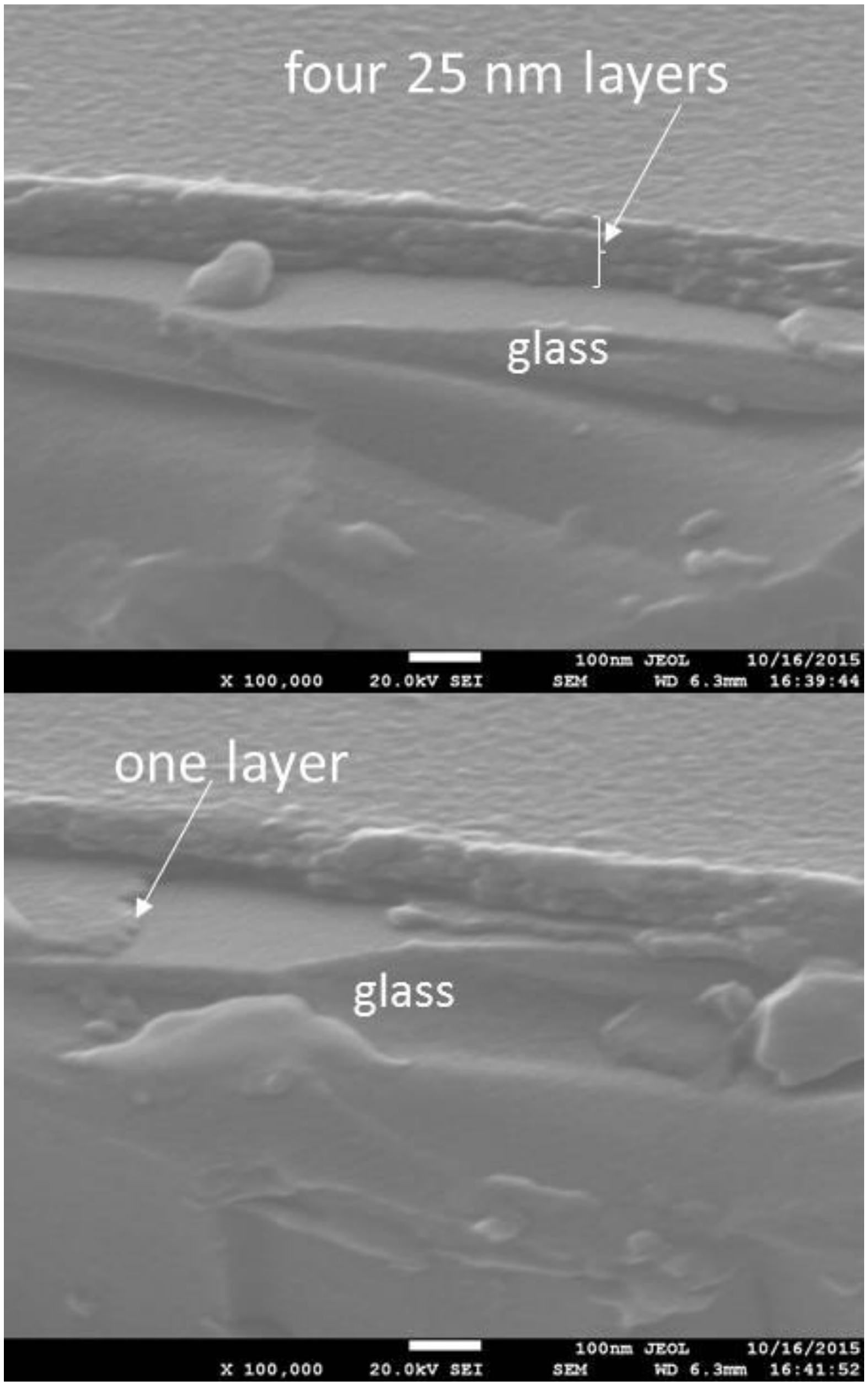

**Fig. 2. Scanning Electron Microscope images of a cleaved four layer Al/$Al_2O_3$ sample.** This sample consists of four 25 nm thick layers of aluminum deposited on top of each other. The top surface of each layer was exposed to air for ~ 1 hr in between the depositions. Upon exposure to the ambient conditions a ~ 2 nm thick $Al_2O_3$ layer was formed on top of each layer. The first 25 nm layer is clearly visible in the bottom image.

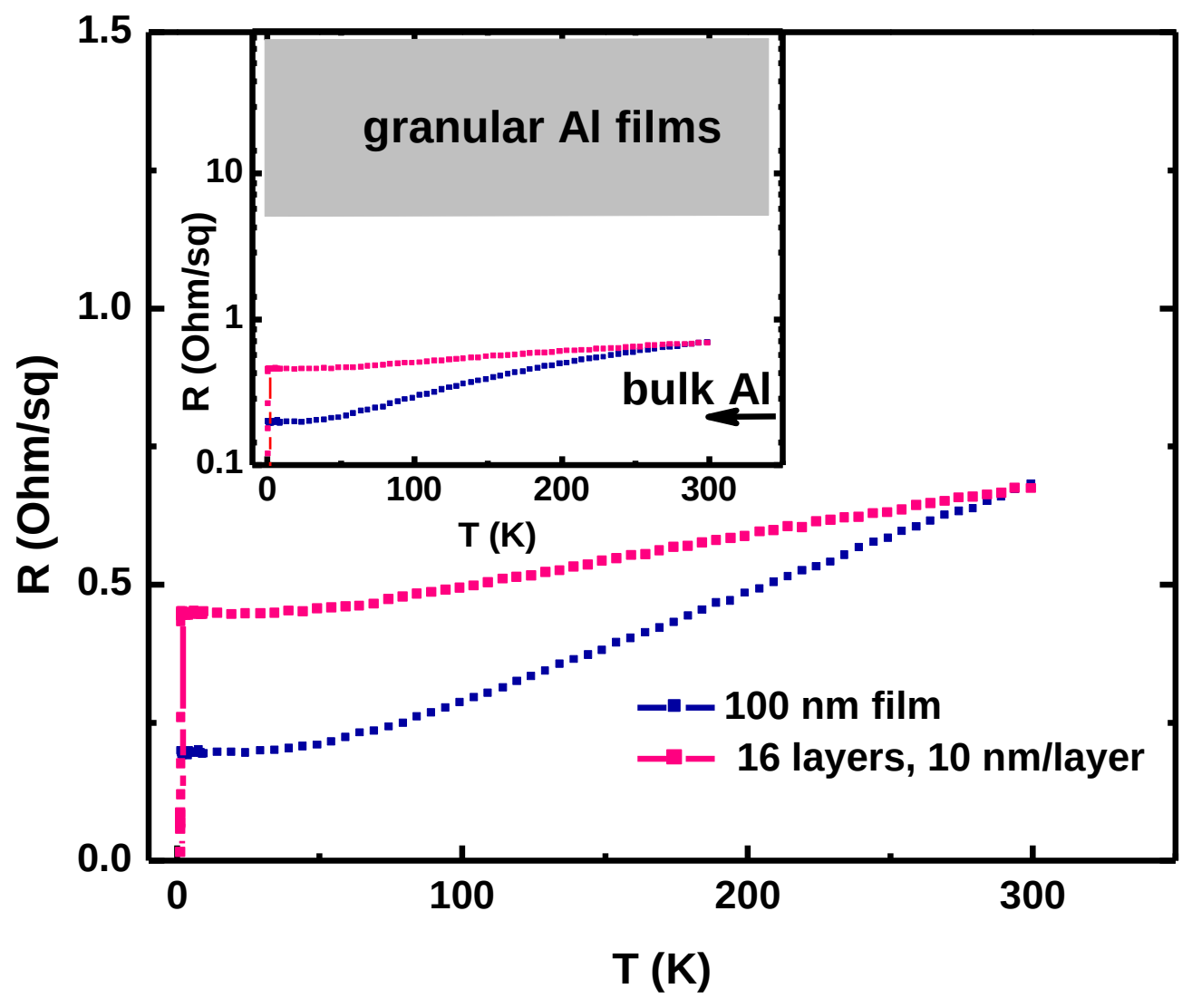

A

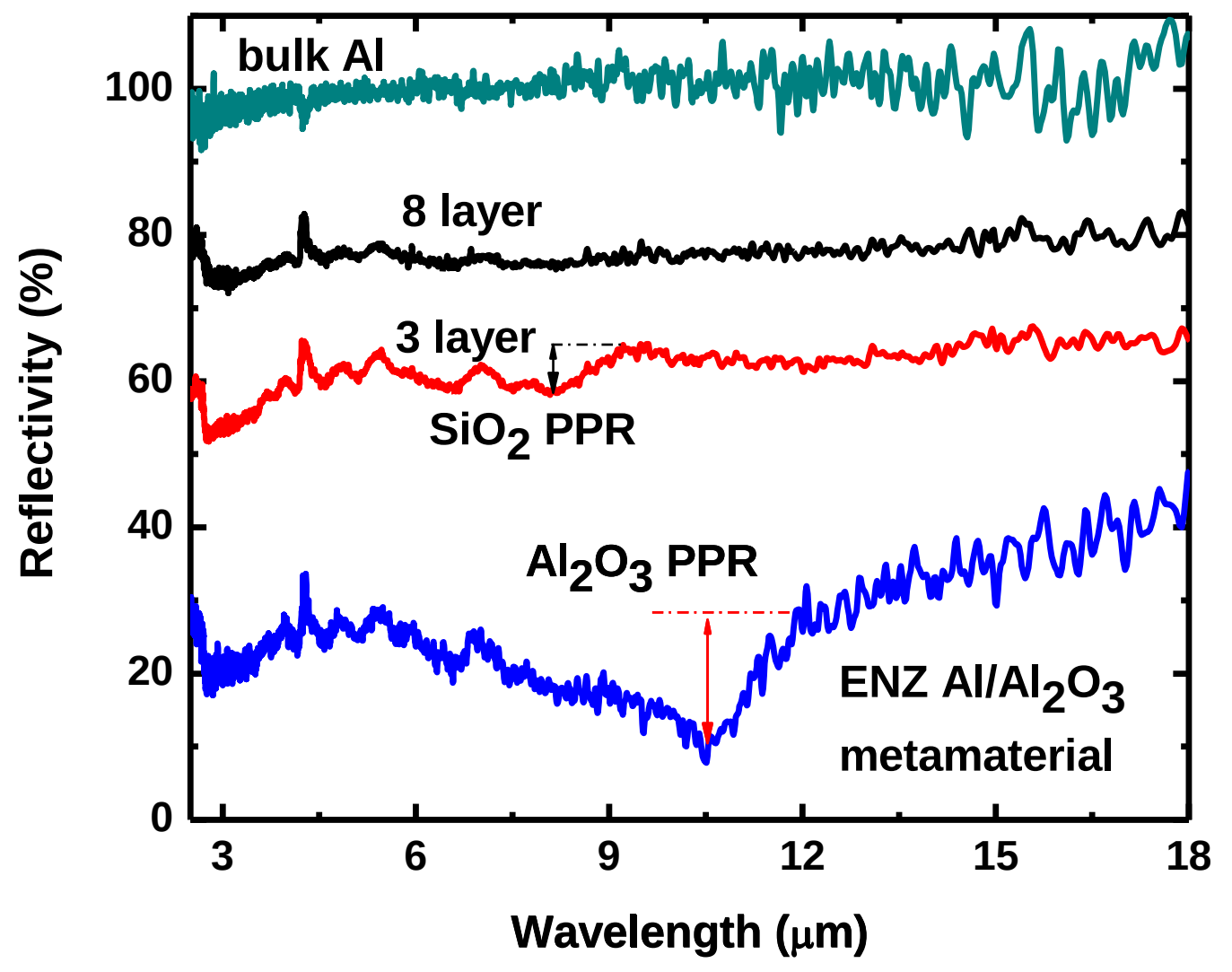

B

**Fig. 3. Measurements of DC and AC (optical) conductivity of the aluminum layers of the fabricated hyperbolic metamaterials.** (**A**) Temperature dependences of the sheet resistance of a 16-layer 10nm/layer Al/$Al_2O_3$ hyperbolic metamaterial and a 100 nm thick Al film. As illustrated in the logarithmic plot in the inset, the conductivity of the metamaterial approaches conductivity values of bulk aluminum and is far removed from the parameter space characteristic for granular Al films which is indicated by the gray area. (**B**) IR reflectivity of bulk aluminium, 3 and 8 layer hyperbolic metamaterial, and the core shell metamaterial samples measured in the long wavelength IR (LWIR) (2.5 - 22.5 μm) range using an FTIR spectrometer. The step in reflectivity around 11 μm is related to the phonon-polariton resonance (PPR) and may be used to characterize the spatial distribution of $Al_2O_3$ in the metamaterial samples.

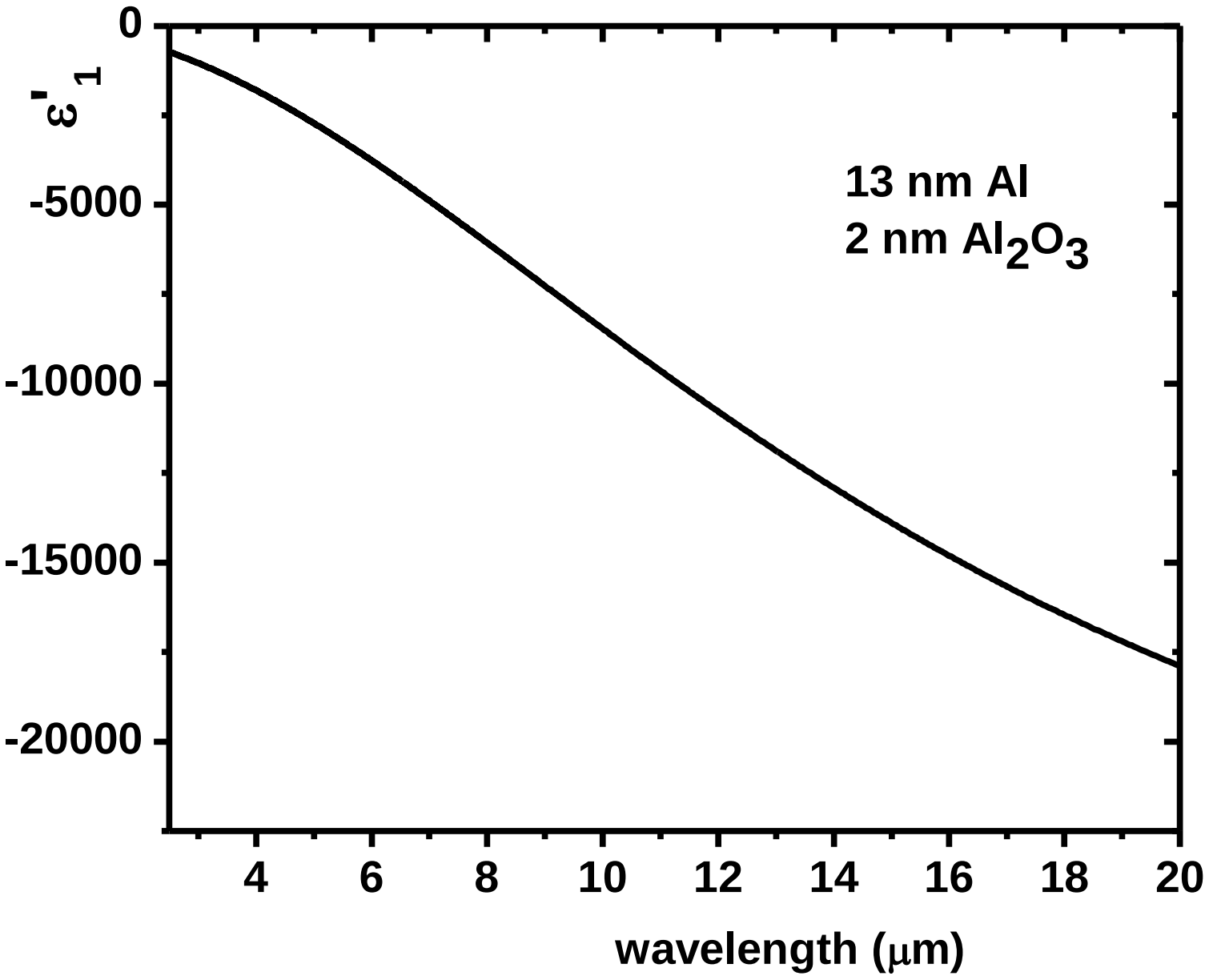

A

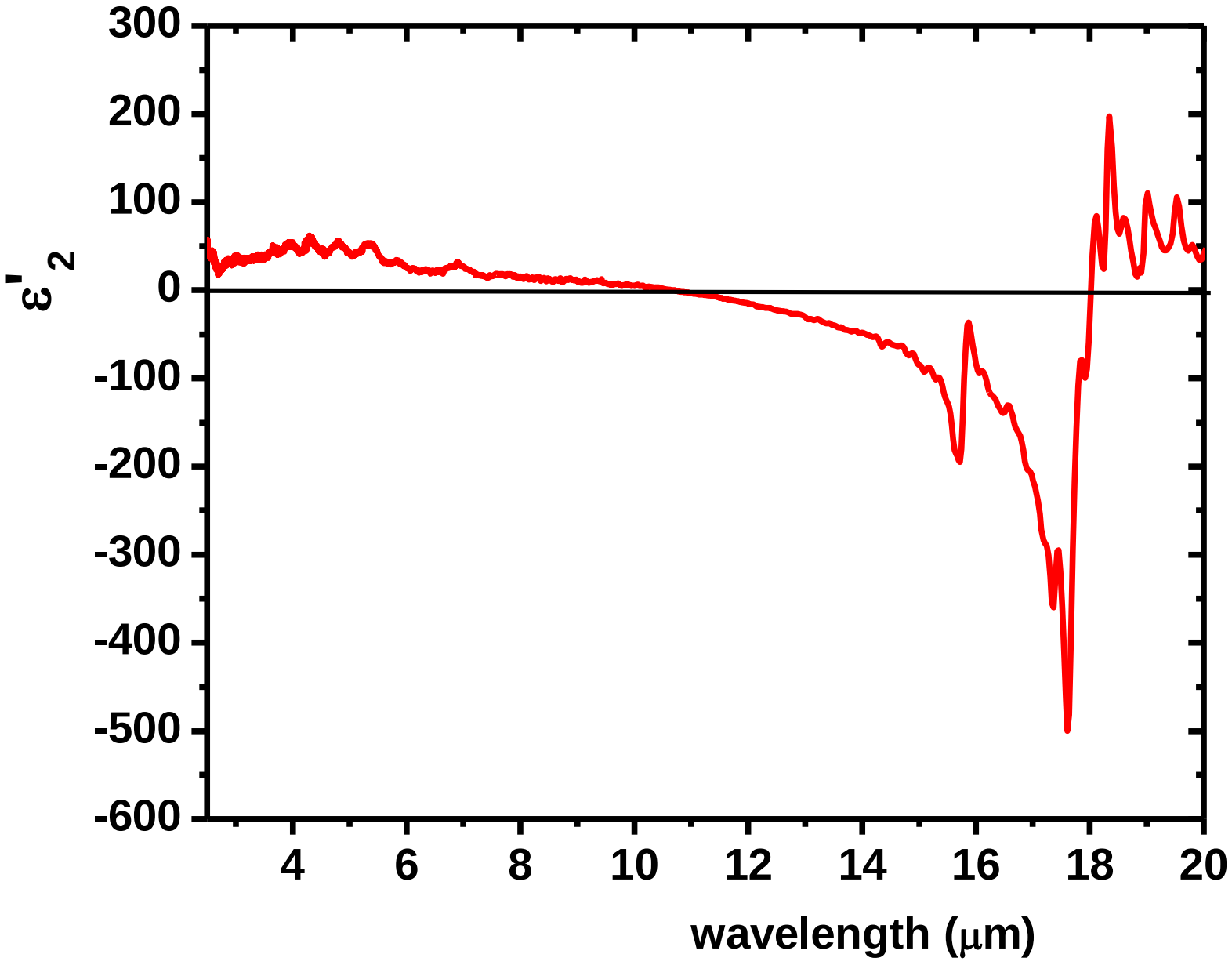

B

**Fig. 4. The calculated plots of the real parts of $\varepsilon_{x,y}$ (A) and $\varepsilon_z$ (B) of the multilayer $Al/Al_2O_3$ metamaterial.** The metamaterial consists of 13 nm thick Al layers separated by 2 nm of $Al_2O_3$ in the LWIR spectral range. The calculations were performed using Eqs. (4,5) based on the Kramers-Kronig analysis of the FTIR reflectivity of Al and $Al_2O_3$ in ref. (4). The metamaterial appears to be hyperbolic except for a narrow LWIR spectral band between 11 and 18 μm.

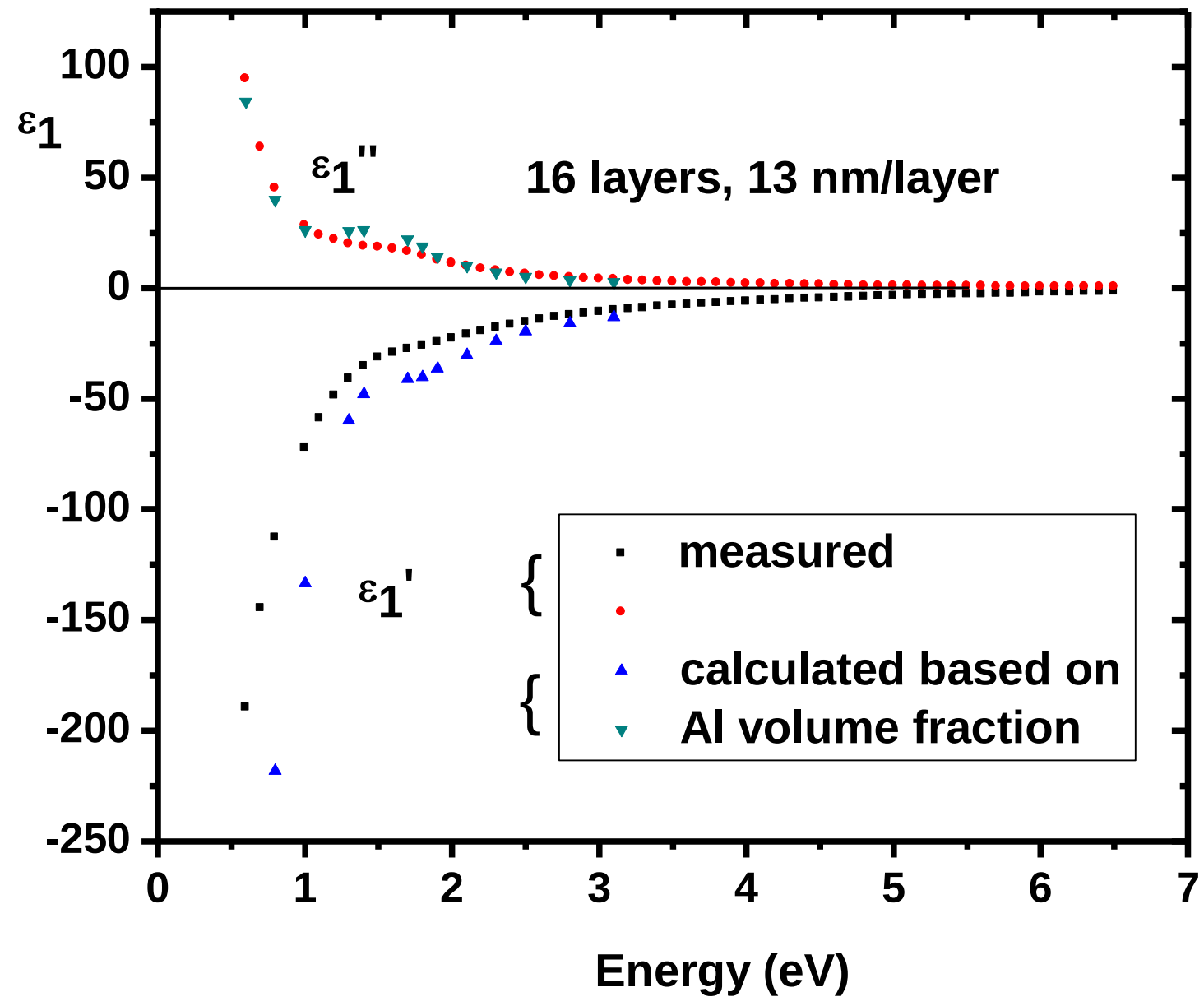

A

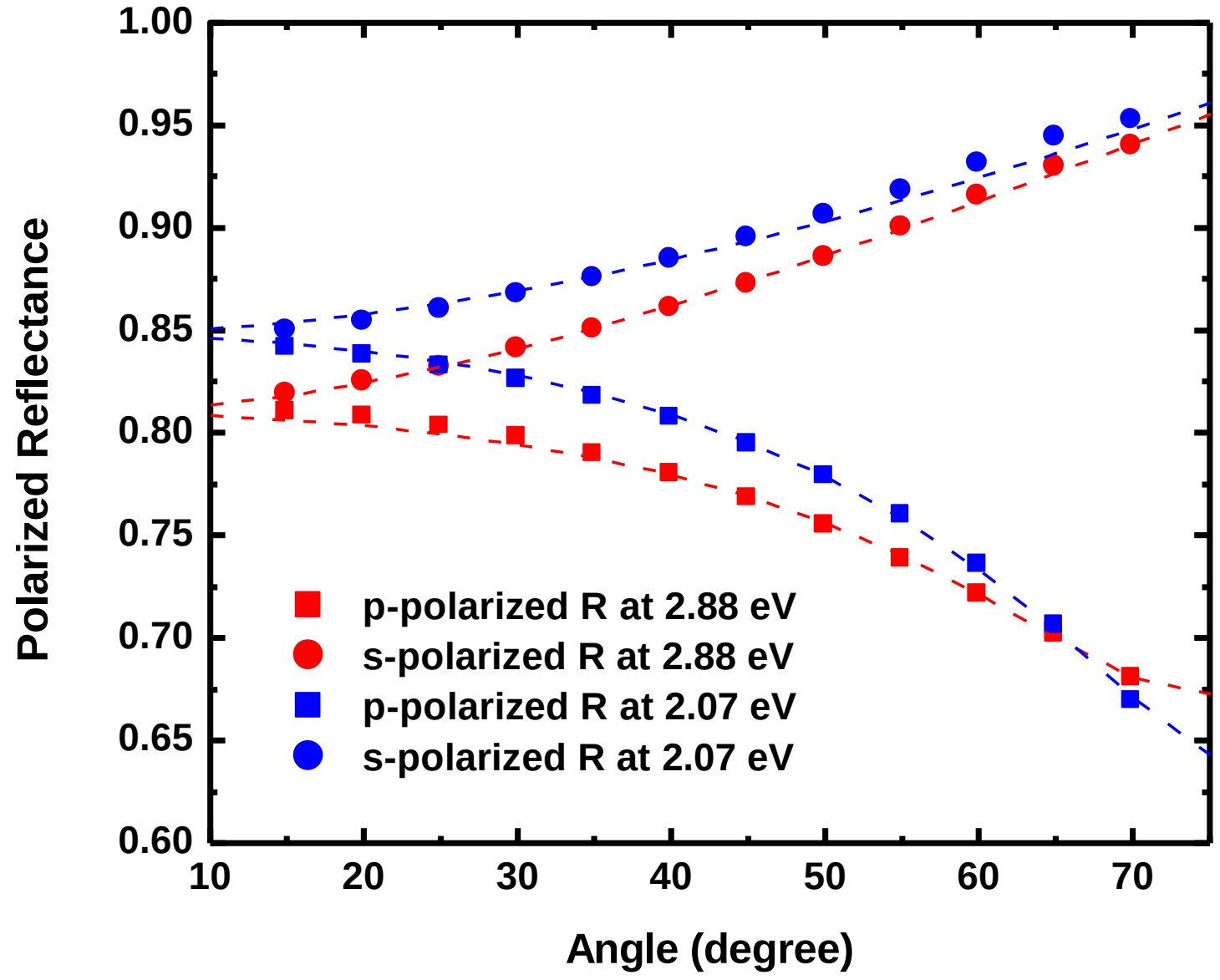

B

**Fig. 5. Ellipsometry and polarization reflectometry of Al/$Al_2O_3$ hyperbolic metamaterials.** (**A**) Comparison of measured pseudo-dielectric function using ellipsometry and theoretically calculated Re$\varepsilon_1$ and Im$\varepsilon_1$. Theoretical data points are based on real and imaginary parts of $\varepsilon_{Al}$ tabulated in (16). (**B**) Data points are measured p- and s-polarized reflectivities of the metamaterial sample at 2.07 eV (600 nm) and 2.88 eV (430 nm). Dashed lines are fits using Eq. (8-11). $\varepsilon_1$ and $\varepsilon_2$ obtained from the fits confirm hyperbolic character of the metamaterial.

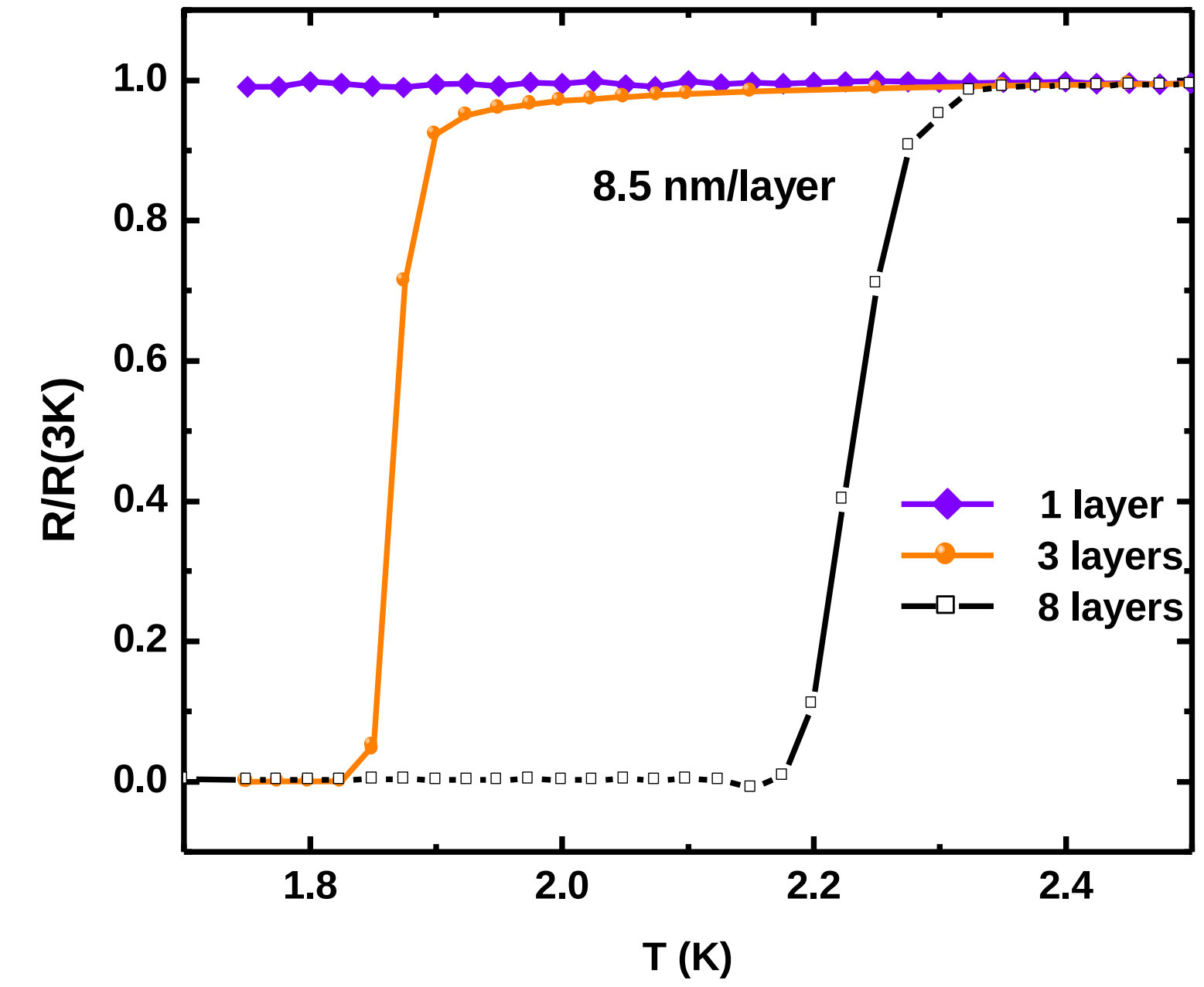

A

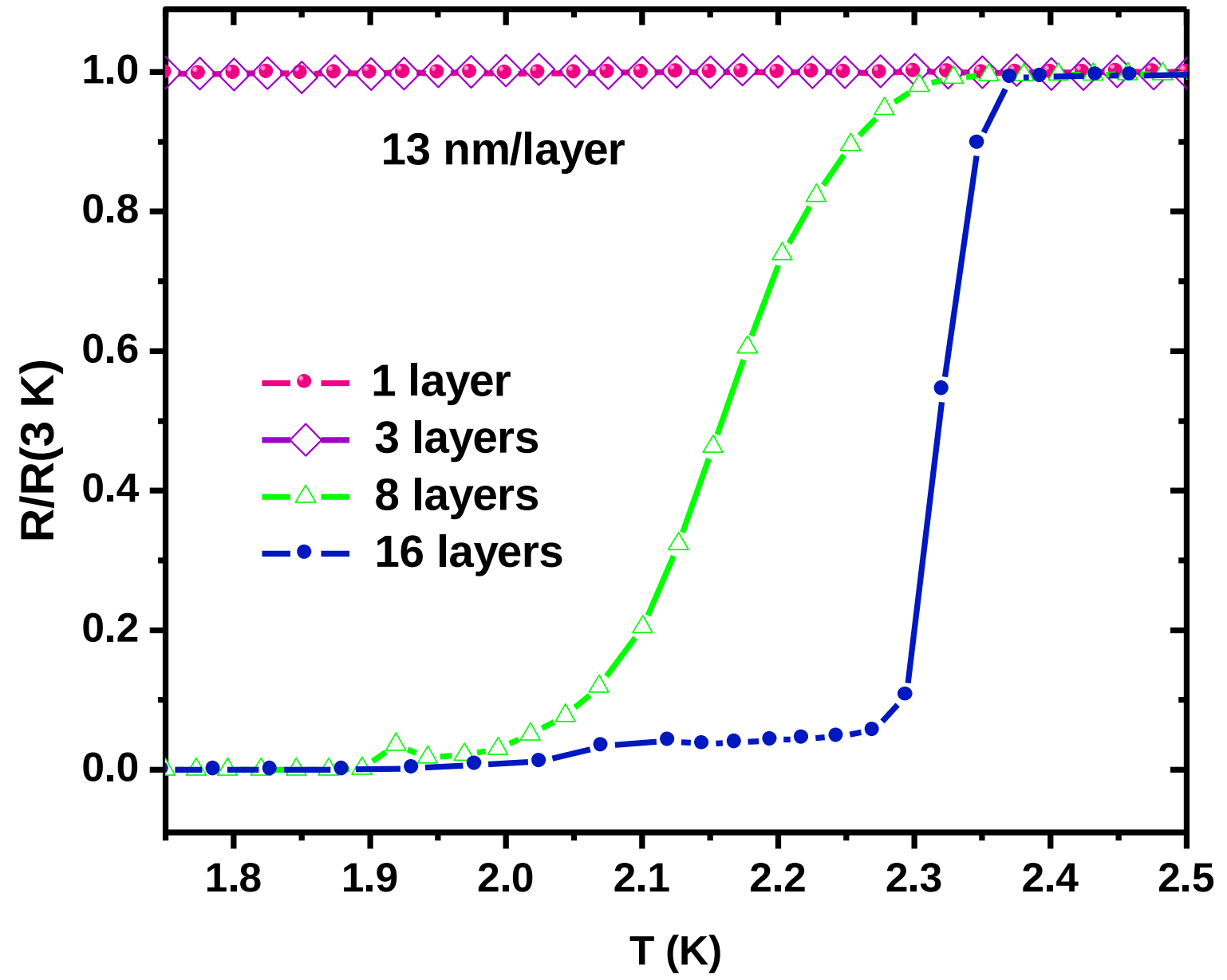

B

**Fig. 6. Effect of the number of layers on $T_c$ of the Al/$Al_2O_3$ hyperbolic metamaterial samples.** (**A**) Measured resistivity as a function of temperature is shown for the 1-layer, 3-layer and 8-layer samples each having the same 8.5 nm layer thickness. (**B**) Measured resistivity as a function of temperature for the 1-layer, 3-layer, 8-layer and 16-layer samples each having the same 13 nm layer thickness.

.

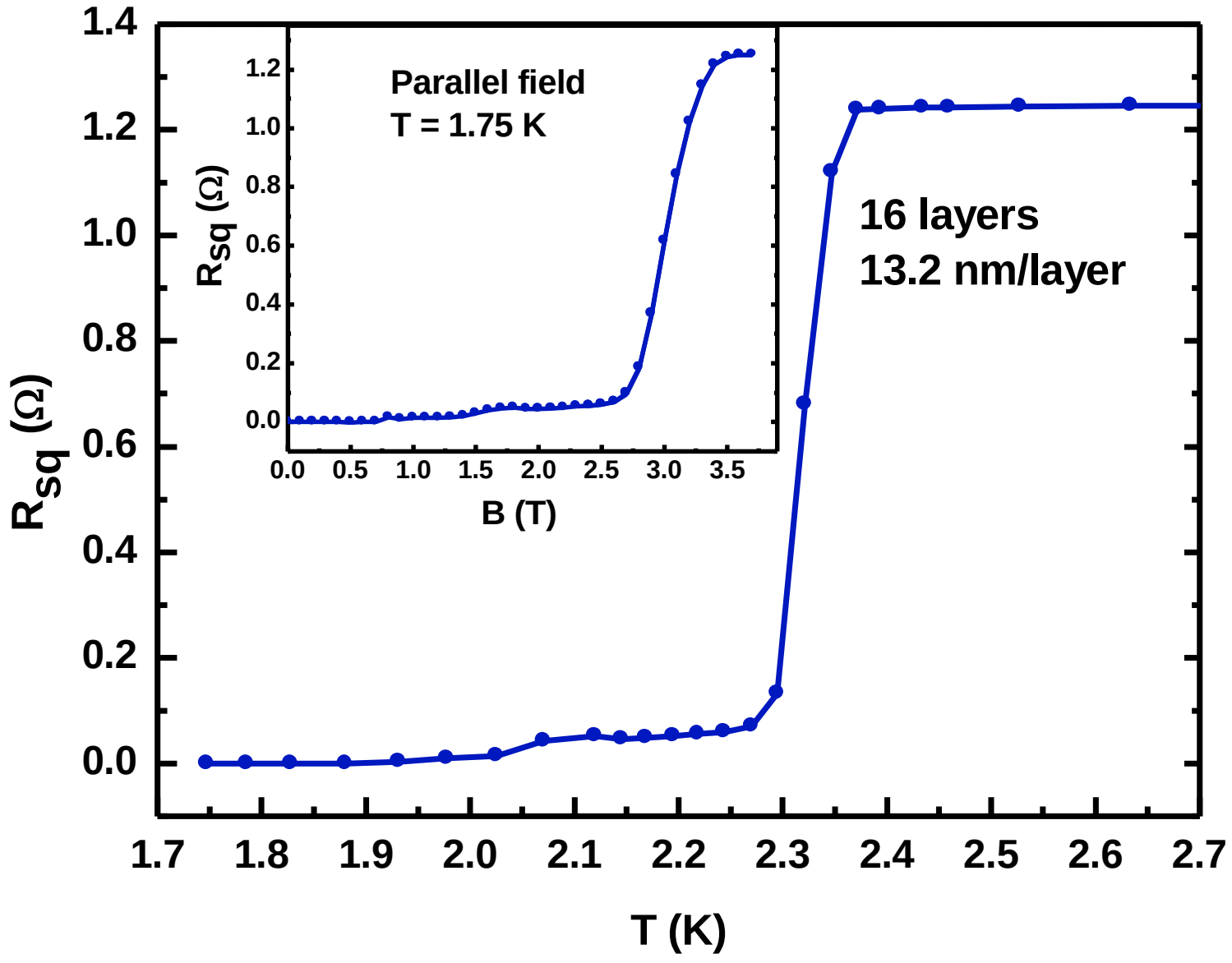

A

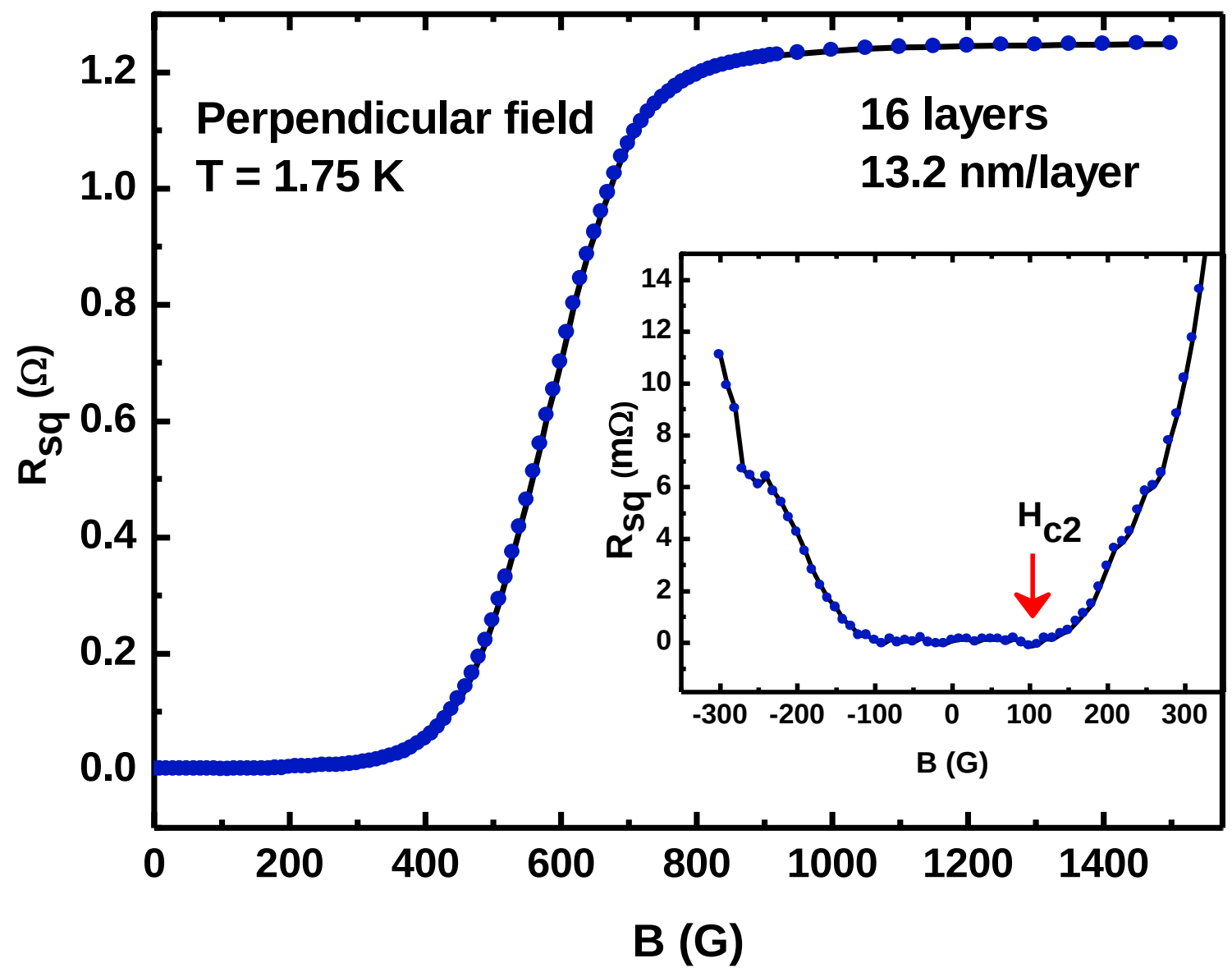

B

**Fig. 7. Evaluation of the Pippard coherence length of the Al/$Al_2O_3$ hyperbolic metamaterial.** (**A**) Measured resistivity as a function of temperature for a 16-layer 13.2 nm layer thickness metamaterial sample. The critical temperature appears to be $T_c$=2.3K. The inset shows resistivity of this sample as a function of parallel magnetic field at T=1.75K. (**B**) Resistivity of the same sample as a function of perpendicular magnetic field at T=1.75K. Assuming $H_{c2}^{perp}$=100G (based on the measurements shown in the inset) the corresponding coherence length appears to be $\xi$=181 nm, which is much larger than the layer periodicity.

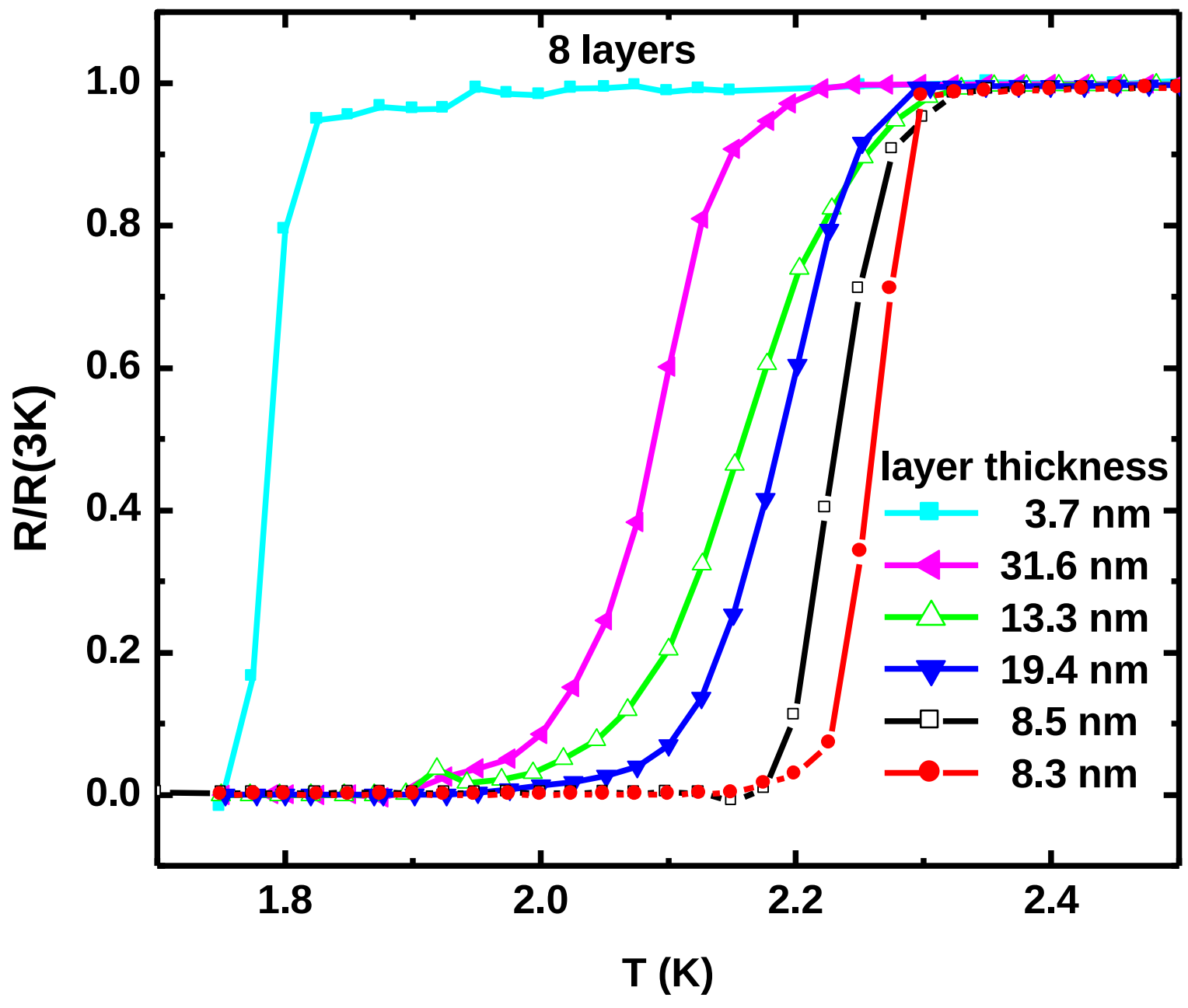

A

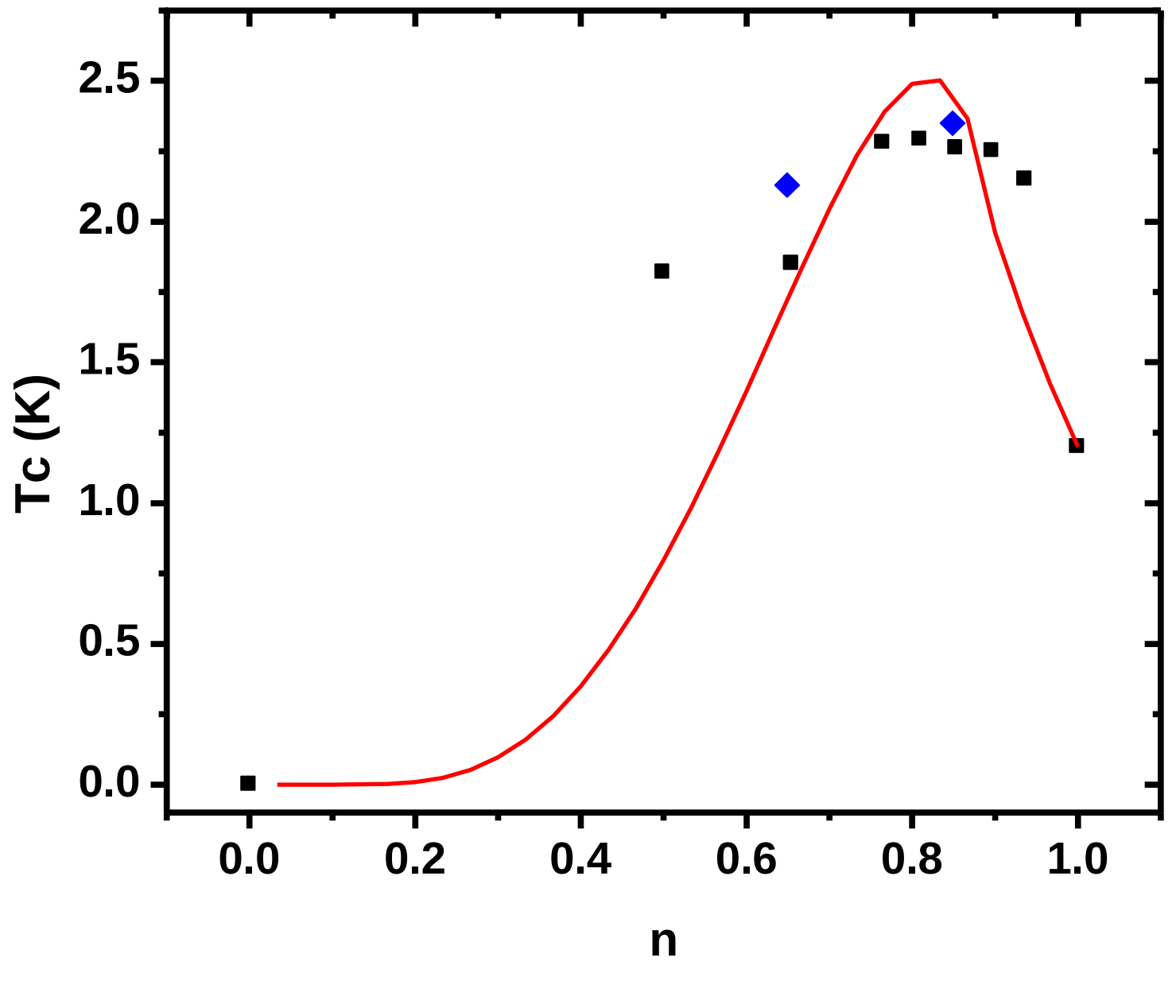

B

**Fig. 8. Effect of the aluminum volume fraction *n* on $T_c$ of the $Al/Al_2O_3$ hyperbolic metamaterial samples.** (**A**) Resistivity as a function of temperature for the 8-layer samples having different aluminum layer thicknesses. (**B**) Experimentally measured behavior of $T_c$ as a function of *n* (which is defined by the Al layer thickness) correlates well with the theoretical fit (red curve) based on the hyperbolic mechanism of $T_c$ enhancement. Experimental data points shown in black correspond to 8-layer samples, while blue ones correspond to 16-layer samples